\documentclass[epsfig,graphics,twocolumn,floatfix,mathbbm,pra,superscriptaddress]{revtex4-1}

\usepackage{graphicx}
\usepackage{amsmath}
\usepackage{amssymb}
\usepackage{gensymb}
\usepackage{braket}
\usepackage{bm} 
\DeclareMathOperator{\sech}{sech}
\usepackage{url}
\usepackage{textcomp}
\usepackage[colorlinks=true]{hyperref}
\usepackage{color}
\usepackage{verbatim}
\usepackage{dcolumn}
\usepackage{booktabs}
\usepackage{multirow}

\newcommand{\nd}[0]{Nd$^{3+}$:Y$_2$SiO$_5$}
\newcommand{\affilge}[0]{Groupe de Physique Appliqu\'ee, Universit\'e de Gen\`eve, CH-1211 Gen\`eve 4, Switzerland}
\newcommand{\affilparis}[0]{Laboratoire Charles Fabry, Institut d'Optique Graduate School, CNRS, Universit\'e Paris-Saclay, 91127 Palaiseau, France}
\newcommand{\affilparistwo}[0]{PSL Research University, Chimie ParisTech, CNRS, Institut de Recherche de Chimie Paris, 75005, Paris, France}
\newcommand{\affilparisthree}[0]{Sorbonne Universités, UPMC Univ Paris 06, 75005, Paris, France}

\begin{document}

\title[a]{Spectral hole lifetimes and spin population relaxation dynamics in \\ neodymium-doped yttrium orthosilicate}%

\author{E.~Zambrini Cruzeiro} 
\author{A~Tiranov} 
\affiliation{\affilge{}}
\author{I.~Usmani} 
\affiliation{\affilparis{}}
\author{C.~Laplane} 
\affiliation{\affilge{}}
\author{J.~Lavoie} 
\altaffiliation[Current address: ]{Present address: Department of Physics and Oregon Center for Optical Molecular \& Quantum Science, University of Oregon, Eugene, OR 97403, USA}
\affiliation{\affilge{}}
\author{A.~Ferrier} 
\affiliation{\affilparistwo{}}
\affiliation{\affilparisthree{}}
\author{P.~Goldner} 
\affiliation{\affilparistwo{}}
\author{N.~Gisin} 
\author{M.~Afzelius} \email[Correspondance: ]{mikael.afzelius@unige.ch}
\affiliation{\affilge{}}

\date{\today}

\begin{abstract}
We present a detailed study of the lifetime of optical spectral holes due to population storage in Zeeman sublevels of Nd$^{3+}$:Y$_2$SiO$_5$. The lifetime is measured as a function of magnetic field strength and orientation, temperature and Nd$^{3+}$ doping concentration. At the lowest temperature of 3 K we find a general trend where the lifetime is short at low field strengths, then increases to a maximum lifetime at a few hundreds of mT, and then finally decays rapidly for high field strengths. This behaviour can be modelled with a relaxation rate dominated by Nd$^{3+}$-Nd$^{3+}$ cross relaxation at low fields and spin lattice relaxation at high magnetic fields. The maximum lifetime depends strongly on both the field strength and orientation, due to the competition between these processes and their different angular dependencies. The cross relaxation limits the maximum lifetime for concentrations as low as 30 ppm of Nd$^{3+}$ ions. By decreasing the concentration to less than 1 ppm we could completely eliminate the cross relaxation, reaching a lifetime of 3.8 s at 3~K. At higher temperatures the spectral hole lifetime is limited by the magnetic-field independent Raman and Orbach processes. In addition we show that the cross relaxation rate can be strongly reduced by creating spectrally large holes of the order of the optical inhomogeneous broadening. Our results are important for the development and design of new rare-earth-ion doped crystals for quantum information processing and narrow-band spectral filtering for biological tissue imaging.
\end{abstract}

\maketitle

\section{INTRODUCTION}

Rare-earth (RE) ions doped into solid-state materials (amorphous or crystalline) are currently investigated in the domain of quantum technology for both storing and processing quantum information \cite{Bussieres2013,RiedmattenAfzeliusChapter2015}. A strong motivation behind this effort is the long optical and spin coherence times that can be achieved at low temperatures~\cite{Sun2002,Zhong2015}. The large number of RE ions that can be considered (RE = Eu, Pr, Tm, Nd, Er, Yb, Ce, etc.) also implies a wide range of possibilities in terms of optical wavelength (ultraviolet to near-infrared), spin transition frequencies (MHz to GHz) and transition dipole moments of the relevant optical and spin transitions \cite{Tittel2010b}.

The RE ions can be grouped into Kramers or non-Kramers ions, depending on the number of 4$f^N$ electrons in the RE$^{3+}$ state of the ion \cite{Tittel2010b}. Kramers ions have an odd number of electrons, while non-Kramers ions have an even number of electrons. In low-symmetry crystallographic sites, the non-Kramers ions have a completely lifted $J$ degeneracy and the ground state spin structure results from nuclear Zeeman and nuclear quadrupole type interactions. These nuclear states generally have long coherence times \cite{Fraval2004,Zhong2015} and can be used as qubits \cite{Fraval2005,Rippe2008} or as long-duration storage states for optical quantum memories \cite{Heinze2013,Gundogan2015,Laplane2016a}. However, the spin transition frequencies are low, in the 10-100 MHz range, which limits the useful bandwidth and the speed with which one can manipulate the spin states. 

In Kramers ions the degeneracy is not completely lifted by the interaction with the crystal lattice. In low-symmetry crystallographic sites, Kramers ions have a two-fold $J$ degeneracy of the ground state (a Kramers doublet). The doublet can often be treated as an effective $S=1/2$ spin with a magnetic moment in the range of 1 - 10 Bohr magnetons $\mu_{\mathrm{B}}$ (in erbium as high as $15\mu_{\mathrm{B}}$). This effective spin model can break down at high magnetic fields and/or low crystal-field splittings \cite{Marino2016}. Using Kramers doublets one can achieve spin transition frequencies in the GHz range by applying a moderate magnetic field, which implies large bandwidth and fast operations. Several Kramers ions also have relevant optical transitions that are easily accessible with diode lasers, such as Nd (883~nm), Yb (980~nm) or Er (1530~nm), an important practical aspect. On the other hand, the large magnetic moments of Kramers ions couple more strongly to lattice phonons and to other magnetic ions in the lattice, which might shorten the spin population ($T_1$) and coherence ($T_2$) times, as well as the optical coherence lifetimes, with respect to non-Kramers ions. To fully exploit the advantages of Kramers ions it is therefore important to understand and optimise their spin properties.

In this article the focus is on the spin relaxation mechanisms of a Kramers ion. The relaxation processes strongly affect the degree of spin polarization that can be achieved through optical pumping, which is a crucial step for quantum processing and storage schemes using both single spins and ensembles of spins. The spin population lifetime also puts an upper limit on the achievable spin coherence time such that $T_2 \leq 2T_1$. A long spin population time is thus a basic requirement for many quantum applications.

Optical pumping and partial spin polarization in Kramers doublets was first observed by Macfarlane and Vial in Nd$^{3+}$:LaF$_3$ \cite{Macfarlane1987}. Specifically, they used spectral hole burning (SHB) to optically pump ions into a Kramers sub-level for a small sub-ensemble of ions in the large optical inhomogeneous broadening. Only recently, however, the optical pumping using Kramers doublets received renewed interest in the context of quantum light storage experiments \cite{Usmani2010,Clausen2011,Zhou2012}. There, efficient optical pumping using SHB is a requirement for achieving high storage efficiencies. This led to a few limited studies of spin population lifetimes measured using SHB. In both neodymium and erbium doped single crystals the lifetimes have been limited to about 100 ms \cite{Hastings-Simon2008a,Afzelius2010,Usmani2010}, which in turn reduces the maximum efficiency of quantum storage protocols \cite{Lauritzen2008}. In general the lifetime limitation is thought to be due to the spin-lattice relaxation (SLR) and/or spin cross relaxation (flip-flop or FF) processes \cite{Bottger2006a,Saglamyurek2015}. But the relative importance of these two processes remains unknown, particularly at the low doping concentrations often used in the context of quantum storage experiments ($<$100 ppm). 

In this article we experimentally characterize the spin population dynamics of a Kramers ion, as a function of the applied magnetic field, temperature and dopant concentration. Specifically, we study neodymium-doped Y$_2$SiO$_5$ crystals, which is a typical Kramers case that we believe is representative of a large class of Kramers-ion doped crystals. One of our main findings is that the spin FF process is limiting the spectral hole lifetime at concentrations as low as 30 ppm. Only in a extremely low-doped sample ($<$ 1 ppm) did we measure a spectral hole lifetime solely limited by the SLR, where the lifetime approaches 4 seconds at low magnetic fields. As the neodymium ion has a moderate magnetic moment among Kramers ions, we expect that the spin FF process has a large impact on the spectral hole lifetime of many Kramers doublets. 

The article is organized as follows. In Sec.~\ref{sec:theory} we discuss the different relaxation processes (SLR and spin FF) and their expected dependence on relevant experimental parameters. We also discuss the difference in measuring the population lifetime using SHB and more conventional electron paramagnetic resonance (EPR) experiments. In Sec.~\ref{sec:exp}, we discuss basic properties of neodymium-doped Y$_2$SiO$_5$ and the employed experimental methods. In Sec.~\ref{sec:results} we present measurements of the spectral hole lifetime as a function of the external magnetic field  (in Sec.~\ref{subsec:field}), temperature (Sec.~\ref{subsec:temp}), neodymium concentration (Sec.~\ref{subsec:concentration}) and overall spin polarization (Sec.~\ref{sec:PIT}). In Sec.~\ref{sec:Conclusions_Outlook} we summarize our results and give an outlook on possible future experiments.

\section{THEORY}
\label{sec:theory}

\subsection{Kramers doublets}
\label{subsec:theory_Kramers_doublets}

\begin{figure}[t!]
 \centering
 \includegraphics[scale=0.42]{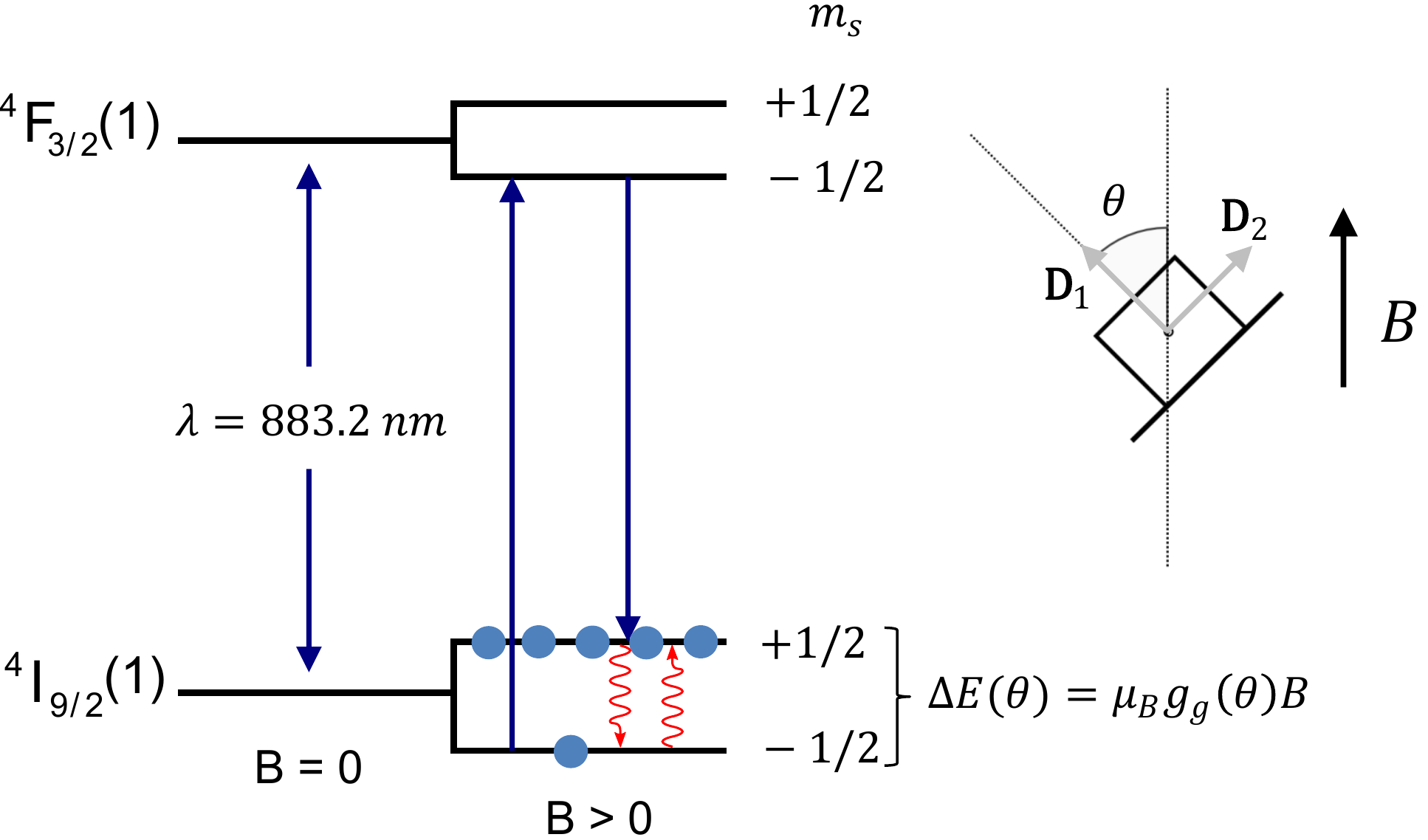}
 \caption{\label{figure1} Kramers doublet splitting under an external magnetic field $B$ for the case of \nd{}. Each crystal field level within the electronic states $^4\mathrm{I}_{9/2}$ and $^4\mathrm{F}_{3/2}$ consists of two magnetic sublevels $m_s=\pm 1/2$. Note that only the lowest crystal field level of each electronic state is shown here. The magnetic sublevels become non-degenerate under the presence of a magnetic field. The splitting in energy depends linearly on the magnetic field $B$ and on the effective $g$ factor $g(\theta)$, which characterizes the angular dependence of the splitting. On the right, the convention used for the magnetic field angle $\theta$ is shown. The magnetic field is static and is applied in the $\textbf{D}_1$-$\textbf{D}_2$ plane, where $\textbf{D}_1$ and $\textbf{D}_2$ are the so-called polarization extinction axes of Y$_2$SiO$_5$ crystal. Absorption is maximal when the polarization of light is linear and aligned with $\textbf{D}_1$.
}
\end{figure}

We here consider the electronic ground state $^{2S+1}L_J$ of a Kramers ion such as neodymium, ytterbium or erbium. If the site symmetry of the ion is sufficiently low, then the interaction with the crystal lattice splits the $2J+1$ magnetic sublevels into $J+1/2$ Kramers doublets. At the low temperatures considered here ($3 \leq T \leq 5.5$~K), only the lowest Kramers doublet is populated. Moreover, only the doublet with the lowest energy typically has long population lifetimes, while the other doublets have very short lifetimes due to fast phonon emission towards the lowest doublet. We note also that the optically excited state is also a Kramers doublet, although its spin dynamics is not characterized in this work. In the following, all experimental data relates to the lowest doublet in the electronic ground state (Fig.~\ref{figure1}).

Under application of a magnetic field each doublet splits into two levels. The doublet can be modelled as a spin-half system $S=1/2$ with a corresponding spin Hamiltonian $\textbf{H}= \mu_B \textbf{B} \cdot \tilde{g} \cdot \textbf{S}$ \cite{Abragam1970}. Here $\mu_B$ is the Bohr magneton, $\textbf{B}$ is the magnetic field vector, $\tilde{g}$ the $g$ factor matrix and $\textbf{S}$ the spin operator vector. The $\tilde{g}$ matrix is often highly anisotropic, which results in a strong angular dependence of the Zeeman energy split $\Delta E$. In this work we vary the magnetic field in a certain plane (see Fig.~\ref{figure1}), such that the energy split can be written in terms of an effective, angle-dependent $g$ factor $g(\theta)$, i.e. $\Delta E (\theta)=\mu_Bg(\theta)B$. 

At a given temperature $T$, the ratio of spins in the two levels is given by the Boltzmann distribution for a system in thermal equilibrium. At $3\ \mathrm{K}$, and for magnetic fields around $1\ \mathrm{T}$, these levels are roughly equally populated, given that $g(\theta)$ varies between 1.5 and 2.7 in the plane of interest (see Fig.~\ref{figure1}) \cite{Wolfowicz2015}. The goal of optical pumping is to create a population distribution far from thermal equilibrium, such as a completely spin-polarized state with all ions in one of the doublet states. After the optical pumping the spins will rethermalize because of the different population relaxation mechanisms. This, in turn, will limit both the time during which the desired state can be used and the maximum degree of spin polarization that can be achieved. We will therefore start by discussing the different relaxation mechanisms that are relevant for Kramers ions at low temperatures.

\subsection{Spin lattice relaxation}
\label{subsec:theory_SLR}

The Kramers doublet states can thermalize to the bath temperature through different interactions with phonons, which together are denoted as spin lattice relaxation (SLR) \cite{Orbach1961,Scott1962,Larson1966,Larson1966a}. The SLR rate is a single-ion property, i.e. it has no dependence on the concentration of paramagnetic ions. In some rarer cases, however, a spin concentration dependence can be observed due to the ``phonon bottleneck'' phenomenon \cite{Scott1962}.

There are three main types of SLR processes; direct, Raman and Orbach. The direct process involves the absorption or emission of a phonon with the same energy as the doublet energy separation $\Delta E (\theta)$. This process is thus strongly dependent on the density of phonons at a given energy, which scales as $\Delta E^2(\theta)$. The Raman and Orbach processes, on the other hand, are two-phonon processes. The Raman process only requires a two-phonon resonance and therefore uses a larger range of the phonon spectrum. The Orbach process is resonantly enhanced by also involving a one-phonon resonance with a second Kramers doublet, with an energy separation $\Delta_\mathrm{O}$ with respect to the ground state doublet. The three processes add up to a total SLR rate that can be written as \cite{Larson1966,Larson1966a},
\begin{align}\label{eq1}
R_{\mathrm{SLR}} &= \alpha_\mathrm{D}(\theta) g^3(\theta) (\mu_\mathrm{B} B)^5 \coth(\frac{\Delta E (\theta)}{2 k_\mathrm{B} T}) \nonumber \\ 
&+ \alpha_\mathrm{R} T^9+\alpha_\mathrm{O} e^{-\frac{\Delta_\mathrm{O}}{k_B T}},
\end{align}
\noindent where $\alpha_\mathrm{D}$, $\alpha_\mathrm{R}$, $\alpha_\mathrm{O}$ are the coupling parameters for the direct, Raman and Orbach processes, respectively.

The Raman and Orbach processes are strongly temperature dependent, but they normally have no magnetic-field dependence. A field dependence might appear, however, if the Zeeman split $\Delta E (\theta)$ becomes comparable to the crystal-field split with respect to the first excited crystal-field level \cite{Larson1966a}. Since all experiments presented in this article are far from this regime we consider the Raman and Orbach processes to be field-insensitive.

Kurkin and Chernov~\cite{Kurkin1972} have measured the Raman and Orbach coupling parameters in Nd$^{3+}$:Y$_2$SiO$_5$ using EPR techniques. They found  $\alpha_\mathrm{R}=1.2 \cdot 10^{-5}$, $\alpha_\mathrm{O}=3.8 \cdot 10^{10}$ and $\Delta_\mathrm{O}/k_B = 97$ K (for the crystallographic site relevant to this article, see Sec.~\ref{subsec:crystal}). Using these parameters we calculate that, for temperatures of 3 and 5 K, the Raman and Orbach processes combined amount to a SLR rate of 0.24 and 166 Hz, respectively. These rates correspond to population lifetimes of 4.2 s and 6 ms, respectively. For efficient optical pumping the population lifetime must be much longer than the radiative lifetime of the optically excited state (see Sec.~\ref{subsec:shb}), which is 300~\textmu s in Nd$^{3+}$:Y$_2$SiO$_5$ \cite{Usmani2010}. One can thus immediately conclude that for Nd$^{3+}$:Y$_2$SiO$_5$ pumping cannot be efficient at a temperature of 5 K or above. Since Raman and Orbach parameters have been measured for many Kramers ions in different host crystals, such a simple analysis permits to evaluate below which temperature efficient optical pumping could potentially be achieved.

The direct process is only weakly dependent on temperature, as compared to Raman and Orbach, but displays a strong dependence on both the angle and magnitude of the magnetic field. As a consequence much less information can be found in the literature, typically the direct contribution to the relaxation rate is characterized only for a fixed angle and magnetic field, such as for Nd$^{3+}$:Y$_2$SiO$_5$~\cite{Kurkin1980}. In the limit where $\Delta E \ll 2 k_\mathrm{B} T$, which holds for most of the data presented in this article, the direct process scales as $\alpha_\mathrm{D}(\theta) g^2(\theta) B^4$. There is thus often a known angular dependence due to $g^2(\theta)$, but the angular dependence in the coupling parameter $\alpha_\mathrm{D}(\theta)$ is generally unknown. One of the goals of this work is to measure $\alpha_\mathrm{D}(\theta)$ in Nd$^{3+}$:Y$_2$SiO$_5$.

\subsection{Cross relaxation}
\label{subsec:theory_FF}

Another type of spin relaxation process is the cross relaxation between two spins \cite{Portis1956,Bloembergen1959,Larson1966a}, which is also called the spin flip-flop (FF) process. If two spins with the same energy splitting $\Delta E$ are spatially close enough they can swap excitation through a magnetic dipole-dipole interaction. As a consequence it depends on the concentration of spins. In EPR experiments the cross relaxation process is often considered between two different types of paramagnetic ions, ensembles A and B, which are tuned into resonance by making their $g(\theta)$ factors similar for specific angles of the magnetic field \citep{Larson1966a}. If ensemble A has been saturated by an initial microwave pulse, its spins can relax by flip-flopping with the ensemble B spins, causing an increased relaxation rate at those specific angles of the magnetic field.

\begin{figure}[t!]
 \centering
 \includegraphics[scale=0.7]{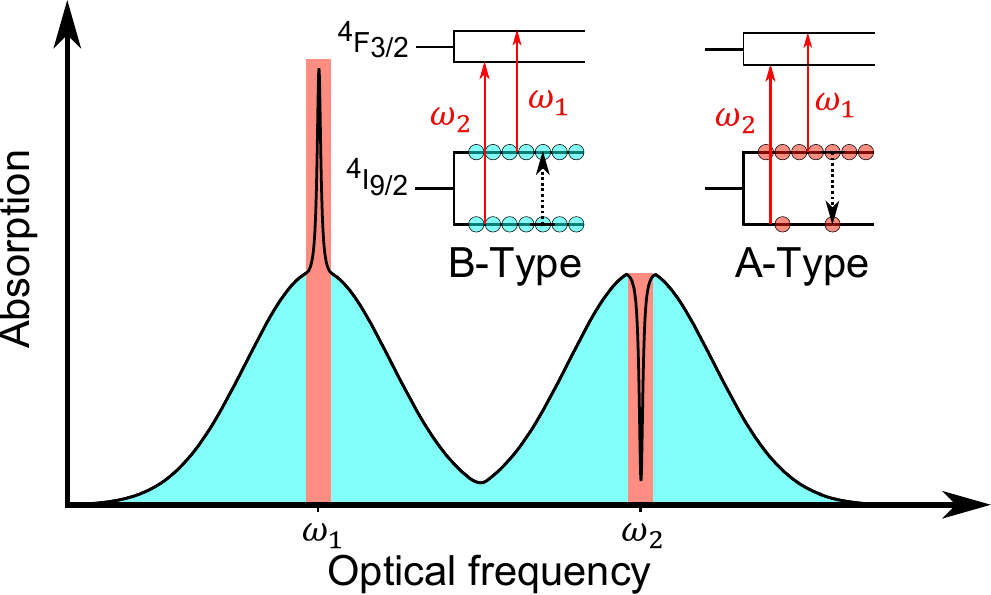}
 \caption{\label{figure2} Illustration of the creation of two spin populations by spectral hole burning on an optical transition within a four-level system as shown in Fig.~\ref{figure1}. For simplicity we assume that only two transitions are allowed, originating from each $^4\mathrm{I}_{9/2}$  spin level. These optical transitions are inhomogeneously broadened with average frequencies $\omega_1$ and $\omega_2$. A narrow hole is burnt into the $\omega_2$ absorption line by optically pumping the spins into the upper $^4\mathrm{I}_{9/2}$ spin level. This creates a highly polarized spin ensemble in that narrow spectral region, denoted ensemble as A (in red), while the remaining ions have a thermal distribution, denoted ensemble as B (in cyan). A spin of ensemble A can flip-flop with a spin of ensemble B, as depicted by dashed arrows. This causes a decrease in the spin polarization of ensemble A and a time-dependent decay of the spectral hole. The associated increase in absorption on the $\omega_1$ line is not measurable since it is distributed over the large spectral region made up of ensemble B spins.}
\end{figure}

In our experiment we perform optical pumping using spectral hole burning and we need to consider how cross-relaxation can affect the lifetime of the spectral hole. The spectral hole burning creates a strongly spin-polarized ensemble A for a small frequency range within the large optical inhomogeneous broadening. All other spins, ensemble B,  remain, however, in a thermal distribution between the two spin states (see Fig.~\ref{figure2}). We stress that ensembles A and B both contain Nd$^{3+}$ ions with identical spin properties, but whose optical frequencies are different (Fig.~\ref{figure2}). In typical spectral hole lifetime measurements~\cite{Hastings-Simon2008,Hastings-Simon2008a}, ensemble A contains much less than $1\%$ of the total number of spins. For some broadband quantum memory applications, this fraction can approach $10\%$ \citep{Bussieres2014,Saglamyurek2011}. We note that the effect of spin flip-flops on the spectral hole lifetime has also been considered in EPR experiments \cite{Bloembergen1959}. There, however, only off-resonant flip-flops can cause a decay of the \textit{spin} hole, while in our case resonant spin flip-flops can cause a decay of the \textit{optical} hole.

In our spin flip-flop model we assume that there is no correlation between the inhomogeneous broadening of the spin transition and the optical transition. Hence, the spectral hole only appears on the optical transition and not on the spin transition where the A and B spins cannot be distinguished in frequency, i.e. in $\Delta E$. As a consequence there are always many spins within ensemble B that can resonantly spin flip-flop with the initially spin-polarized ensemble A, effectively causing a fast relaxation and a limitation in the initial spin polarization that can be achieved for ensemble A. This is true as long as $\Delta E \ll 2 k_\mathrm{B} T$, while in the opposite limit a very different behaviour can be expected depending on if spins A are polarized into the upper or lower energy level. 

The calculation of the cross-relaxation rate between identical spins having isotropic $g$ tensors was first considered by Portis~\cite{Portis1956}, in the limit of equally populated spin states ($\Delta E \ll 2 k_\mathrm{B} T$). B\"{o}ttger et al. \cite{Bottger2006a} proposed a modified formula which is valid for any temperature range, which can be written as,
\begin{equation}\label{eq2}
R_{\mathrm{FF}}=\alpha_{\mathrm{FF}}\frac{g^4n^2}{\Gamma}\sech^2\left(\frac{\Delta E (\theta)}{2 k_\mathrm{B} T}\right),
\end{equation}
\noindent where $\alpha_{\mathrm{FF}}$ is the coupling parameter, $\Gamma$ is the inhomogeneous spin linewidth and $n$ is the concentration of dopant ions. 

In the case of spins with anisotropic $g$ tensors the calculations are much more complicated \cite{VanVleck1948}. In the Supplementary Material we use a simpler approach to calculate the expected angular dependence of the average spin flip-flop rate. It is based on a dipole-dipole interaction and Fermi's golden rule.  Our calculations show that one can not simply replace the isotropic $g$ factor in Eq.~(\ref{eq2}) with the effective $g(\theta)$ factor for anisotropic spins. In some cases one can still derive simple formulas for the angular dependence, such as for $g$ tensors with axial symmetry and measurements in planes containing the principal axes of the $g$ tensor. In general, however, the angular dependence of the rate cannot be written as a simple formula. Therefore we henceforth write the FF rate as 
\begin{equation}\label{eq2b}
R_{\mathrm{FF}}=\beta_{\mathrm{FF}}(\theta)\frac{n^2}{\Gamma}\sech^2\left(\frac{\Delta E (\theta)}{2 k_\mathrm{B} T}\right),
\end{equation}
\noindent where the angular dependence is in the parameter $\beta_{\mathrm{FF}}(\theta)$.

We further note that for the range of fields and temperatures considered here, the $\sech^2(\Delta E /2 k_\mathrm{B} T )$ term is close to 1, such that one would expect a very weak $B$-field dependence of the FF rate. But on the contrary we will show experimental evidence of a strong $B$ dependence of the FF process, prompting a modification of Eq.~(\ref{eq2b}). Similar results were recently observed in an erbium-doped silicate fiber~\cite{Saglamyurek2015} and we will discuss the similarities and differences with respect to that work in relation to the experimental results in Sec.~\ref{subsec:field}.

\section{EXPERIMENTAL DETAILS}
\label{sec:exp}

\subsection{Nd$^{3+}$-doped Y$_2$SiO$_5$}
\label{subsec:crystal}

The Y$_2$SiO$_5$ host crystal is interesting because its low nuclear spin density generally results in long optical and hyperfine coherence times when doped with RE ions~\cite{Equall1994,Fraval2004,Bottger2006a,Zhong2015}. The optical properties of Nd$^{3+}$:Y$_2$SiO$_5$ have been studied since the mid 1980s for its use as a laser medium~\cite{Tkachuk1986,Beach1990,Beach1990a,Comaskey1993}. The Nd$^{3+}$:Y$_2$SiO$_5$ crystal was first introduced in the field of quantum information with the demonstration of storage of light at the single-photon level in 2010~\cite{Usmani2010}. Since then it has been used in numerous quantum storage experiments~\cite{Clausen2011,Usmani2012,Bussieres2014,Tiranov2015a,Tiranov2016a} and in coherent storage of microwave excitations~\cite{Wolfowicz2015}. A nanophotonic cavity has also been fabricated in a Nd$^{3+}$:Y$_2$SiO$_5$ crystal, showing enhanced interaction between Nd$^{3+}$ ions and light~\cite{Zhong2015a}.

The optical transition of interest for coherent light-matter interactions is between the lowest Kramers doublet in the electronic ground state $^4\mathrm{I}_{9/2}$ and in the electronically excited state $^4\mathrm{F}_{3/2}$. The radiative lifetime of the excited state is about 300 \textmu s~\cite{Usmani2010}, one of the shortest of any RE ion with an optical transition having good coherence properties. The short lifetime makes optical pumping more efficient, given a fixed spin population lifetime, as compared to RE ions having long radiative lifetimes such as erbium (about 10~ms).

The Nd$^{3+}$ ions replace Y$^{3+}$ ions in two possible crystallographic sites in the lattice \cite{Kurkin1980}, both having a site symmetry of $C_1$. In this work only ions in site 1 are studied, following the site notation of Ref.~\cite{Beach1990}, since the corresponding absorption coefficient is higher. The transition wavelength for site 1 is 883.0 nm (11325 cm$^{-1}$). We note that this site notation is inverted with respect to the EPR notation introduced in Ref.~\cite{Kurkin1980}.

All crystals we use have a natural abundance of Nd$^{3+}$ isotopes, hence $80\%$ with zero nuclear spin $I=0$ (with even atomic mass number) and $20\%$ with nuclear spin $I=7/2$ ($12.2\%\ ^{143}\mathrm{Nd}$ and $8.3\%\ ^{145}\mathrm{Nd}$). For the isotopes with $I=7/2$ the ground state has more than the two levels of the simple Kramers doublet. The coupling to the nuclear spin also affects the population lifetime and opens up more decay channels. In this work we aimed at only characterizing even isotopes with $I=0$. In Sec.~\ref{subsec:shb} we discuss how we could extract lifetimes that, to a high confidence, only pertain to even isotopes.

Part of this work concerns the concentration dependence of the spectral hole lifetime, therefore we study three crystals with different concentrations. We use a crystal nominally doped with $30\ \mathrm{ppm}$ Nd$^{3+}$ grown by Scientific Materials and another one with $75\ \mathrm{ppm}$ grown by us. The latter crystal was grown by the Czochralski method using an inductively heated irridium crucible. The starting oxides were of at least $99.99\%$ purity (Alfa Aesar). We also used a crystal more strongly doped with $\mathrm{Eu}^{3+}$ ($1000\ \mathrm{ppm}$), which contains a residual concentration of $\mathrm{Nd}^{3+}$. Using absorption spectroscopy we estimated the $\mathrm{Nd}^{3+}$ concentration to be $0.5\pm 0.3\ \mathrm{ppm}$. Since we expect no cross relaxation between $\mathrm{Nd}^{3+}$ and $\mathrm{Eu}^{3+}$ ions this crystal represents in a good approximation a pure Nd$^{3+}$:Y$_2$SiO$_5$ crystal with extremely low Nd$^{3+}$ concentration.

All crystals were cut along the polarization extinction axes $\textbf{D}_1$,$\textbf{D}_2$ and $\textbf{b}$ \cite{Li1992}, where $\textbf{b}$ coincides with the unit cell axis that has $C_2$ symmetry. Light was propagating along $\textbf{b}$ with its polarization along $\textbf{D}_1$ for highest absorption. The crystals had peak absorption coefficients of $\alpha_{30}= 3\ \mathrm{cm^{-1}}$~\cite{Usmani2010}, $\alpha_{75}= 7.4\ \mathrm{cm^{-1}}$~\cite{Bussieres2014} and $\alpha_{0.5}=0.05\ \mathrm{cm^{-1}}$ (at zero magnetic field). The magnetic field was applied in the $\textbf{D}_1$-$\textbf{D}_2$ plane where all ions are magnetically equivalent. Out of this plane the ions split into two magnetically non-equivalent sub-ensembles, related by the $C_2$ symmetry around $\textbf{b}$, having different Zeeman splittings $\Delta E (\theta)$. This would complicate the interpretation of the data, since these are expected to have different spin population lifetimes.

\subsection{Spectral hole decay measurements}
\label{subsec:shb}

Spectral hole burning (SHB) is a common technique to measure spin dynamics of RE ions at low temperatures~\cite{Koenz2003,Ohlsson2003,Hastings-Simon2008a,Hastings-Simon2008,Saglamyurek2015}. It consists of a burn pulse that optically pumps ions out of a specific ground state, either to the optically excited state or to another ground state through spontaneous emission. By pumping for a long time with respect to the excited state lifetime $T_1^{\mathrm{opt}}$ one can polarize most of the spins in a particular ground state, as shown in Fig.~\ref{figure1}. This requires that the spectral hole lifetime $T_1^{\mathrm{SHB}}=1/(R_{\mathrm{FF}}+R_{\mathrm{SLR}})$ is much longer than $T_1^{\mathrm{opt}}$ and that the branching ratio of the two optical transitions is high enough as discussed thoroughly in Refs~\cite{Louchet2007,Lauritzen2008,Afzelius2010b}.

To probe the hole a second optical pulse measures the population of the optically pumped state after some delay, which allows one to measure the recovery of the thermal population. In our case the shortest delay was much longer than $T_1^{\mathrm{opt}}$, such that the recovery only involved spin dynamics. We also emphasize that we measure the spectral hole area, such that we are not sensitive to spectral diffusion which can decrease the hole amplitude through a time-dependent broadening of the hole. In general we did not observe spectral diffusion in these measurements. Therefore, measuring the decay of the hole area or the hole depth is equivalent in our case.

For all the measurements presented in this article we observed recovery signals consisting of a short decay on the scale of 100 ms or less and a long decay of a few seconds. The hole depth related to the short decay depends strongly on the magnetic field strength, for a fixed angle, and there is a clear correlation with the lifetime given by the decay constant. For some fields this depth corresponded to almost the entire optical depth, which means that it cannot stem from odd isotopes with nuclear spin $I=7/2$ as these only make up 20\% of the ensemble (and hence the total optical depth). In general we can be certain that holes deeper than about 20\% cannot come from odd isotopes. We therefore assume that the short decays with large amplitudes stem from even isotopes with no nuclear spin $I=0$. Each data set for a fixed field angle was examined in this way and only the decay constants of sufficiently deep holes were selected for the final analysis. This limited our measurements to a certain magnetic field range, depending on the crystal (i.e. doping concentration) and the field angle $\theta$.

The long decay is related to a much smaller relative fraction of the hole depth which is consistent with a contribution to the hole from odd isotopes with $I=7/2$. The related hole depth is much less dependent on the magnetic field strength and angle, which is also consistent with its much longer lifetime that would make optical pumping efficient for any field configuration. We have also performed a few SHB measurements on an isotopically enriched $^{145}$Nd$^{3+}$:Y$_2$SiO$_5$ crystal, supporting this hypothesis. The characterization of hyperfine population lifetimes is out of the scope of this article.

\subsection{Experimental setup}
\label{subsec:setup}

The crystal is placed in an optical cryostat which can be cooled down to 3~K. A superconducting magnet mounted inside can produce a variable magnetic field between 0 and 2 T. The crystal is rotated with respect to the magnetic field using a piezo stage. In the center of the rotator there is a small 1.5~mm hole such that a laser beam can pass through the crystal and rotator. The 883.0 nm laser beam is derived from a continuous-wave external cavity diode laser. An acousto-optic modulator (AOM) is used to modulate the intensity and frequency of light. The AOM was used to create the burn and probe pulses, and to scan the probe pulse a few tens of MHz around the spectral hole. A digital-delay generator creates all the trigger signals for the experiment, while an arbitrary function generator drives the AOM. 

\section{RESULTS AND DISCUSSION}
\label{sec:results}

\subsection{Magnetic field dependence}
\label{subsec:field}

\begin{figure}[t!]
 \centering
 \includegraphics[width=0.5\textwidth]{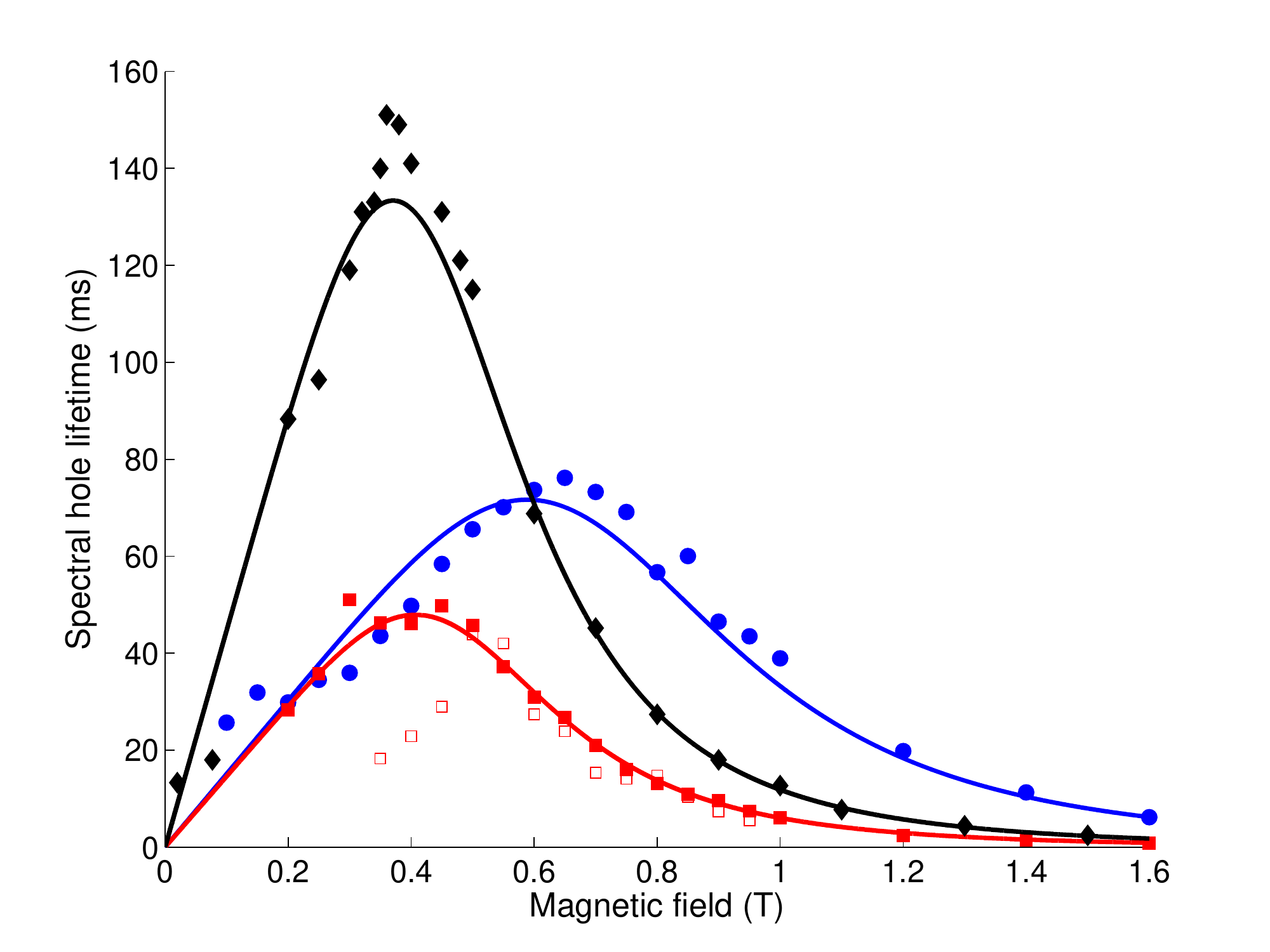}
 \caption{Lifetime of the spectral hole versus magnetic field strength, for a $\text{Nd}^{3+}$ concentration of $30\ \mathrm{ppm}$ and at three different angles $\theta=0\degree$ (blue circles), $\theta=90\degree$ (red filled squares), and $\theta=120\degree$ (black diamonds). Data for $\theta=90\degree$ and the crystal with $75\ \mathrm{ppm}$ $\text{Nd}^{3+}$ concentration is also shown (red open squares). The figure shows both experimental data and fits from the model based on SLR and~FF (see Sec.~\ref{subsec:field} for details). The temperature was 3 K for all data sets.}
 \label{figure_results1}
\end{figure}

In a first series of measurements we study the spectral hole lifetime $T_1^{\mathrm{SHB}}$ as a function of the magnetic field strength for eight fixed angles. The Nd$^{3+}$ concentration was 30 ppm and the crystal was cooled to the temperature of 3~K. Three examples of experimental data sets are shown in Fig.~\ref{figure_results1} for $\theta=0\degree$ ($\textbf{D}_1$ axis), $\theta=90\degree$ ($\textbf{D}_2$ axis) and $\theta=120\degree$.

All measurements display the same general trend as a function of field strength. At low fields the lifetime is small, typically below $10\ \mathrm{ms}$. By increasing the magnetic field one can reach the maximal lifetime at a certain point in the range 0.3-0.6 T, depending on the angle, after which it starts to decrease again. Note that in Fig.~\ref{figure_results1} we show the data that was retained for the final fit to the model, using the procedure discussed in Sec.~\ref{subsec:shb}, but the general trend of a reduction in hole lifetime with decreasing field continues towards zero field. For lower fields, however, it becomes increasingly difficult to use our method for extracting lifetimes of even isotopes with respect to odd isotopes (see Sec.~\ref{subsec:shb}).

The decrease in spectral hole lifetime in the high-field limit is well understood. It is due to the increase in the SLR rate caused by the direct phonon process. As discussed in Sec.~\ref{subsec:theory_SLR} the spectral hole lifetime is expected to scale as  $T_1^{\mathrm{SHB}} \propto 1/(\alpha_\mathrm{D}(\theta) g^2(\theta) B^4)$ in this region. All experimental data sets fits well to the SLR theory in the high-field limit, as shown in Fig.~\ref{figure_results1}.

In the low-field limit the reduction in lifetime cannot be explained by the cross relaxation rate  given by Eq.~\eqref{eq2b}. Indeed, as already discussed in Sec.~\ref{subsec:theory_FF} the $B$-field dependence should be weak, because the thermal population distribution does not change significantly over the relevant range of magnetic fields at 3 K. Yet, by comparing the lifetimes measured for the samples doped with 30 and 75 ppm of Nd$^{3+}$ at the angle of $\theta=90\degree$ ($\textbf{D}_2$ axis), see Fig.~\ref{figure_results1}, it is clear that the $T_1^{\mathrm{SHB}}$ is concentration dependent in the low-field region. Further data showing the concentration dependence are presented in Figs \ref{figure_results4} and \ref{figure7}. There is thus a strong indication of cross-relaxation being the dominant process at low fields.

A similar trend was observed in recent measurements of the spectral hole lifetime in erbium-doped silica glass fibers \cite{Saglamyurek2015}. There it was proposed that the inhomogeneous spin linewidth increases linearly with the magnetic field, which results in an inverse dependence on the field for the cross relaxation rate $R_{\mathrm{FF}} \propto 1/\Gamma \propto 1/(\Gamma_0+\kappa B)$, cf. Eq.~(\ref{eq2b}), where $\kappa$ is a constant and $\Gamma_0$ is the inhomogeneous spin linewidth at zero field. This assumption is justified for an amorphous glass where the inhomogeneous spin linewidth stems from the anisotropy of the $g$-tensor, which generally is large for Kramers ions. In our case a similar linear dependence of the spin linewidth could arise from an inhomogeneity in the $g$ factor, caused by strain or defects \cite{STONEHAM1969}.

There is some experimental support for an increase in the spin linewidth with increasing magnetic field. We performed optically-detected (OD) EPR in the sample with 30 ppm Nd$^{3+}$ concentration, using a method we recently presented in Ref.~\cite{Laplane2016}, which gave a spin linewidth of 5 MHz for a low field ($<10\ \mathrm{mT}$) along $\textbf{D}_2$. In Ref.~\cite{Wolfowicz2015} conventional EPR gave a linewidth of 12~MHz for 561.5~mT along $\textbf{D}_1$, in a sample with 10 ppm Nd$^{3+}$ concentration. But due to the difference in field direction and Nd$^{3+}$ concentration used in the two experiments we cannot make any quantitative estimations of the increase in linewidth. We also note that a field-dependent spin linewidth has been observed in Er$^{3+}$:Y$_2$SiO$_5$ \cite{Probst2013}. Even more recent EPR measurements of the angular dependence of the spin linewidth in Er$^{3+}$:Y$_2$SiO$_5$ also support our hypothesis of a strain-induced field-dependent spin linewidth \cite{Welinski2016b}.

For a spin linewidth of the form $\Gamma = \Gamma_0+\kappa B$ we would expect a weak dependence of the FF rate with the field for sufficiently low fields, hence the $T_1^{\mathrm{SHB}}$ would reach a plateau for low fields. Since we do not observe this (see Fig.~\ref{figure_results1}) we cannot fit $\Gamma_0$ using our data. By using $\Gamma = \kappa B$ as a model, Eq.~(\ref{eq2b}) can then be written as
\begin{equation}\label{eq3}
R_{\mathrm{FF}}=\frac{\gamma_{\mathrm{FF}}(\theta)}{B}\sech^2\left(\frac{\Delta E (\theta)}{2 k_\mathrm{B} T}\right)
\end{equation}
\noindent where $\beta_{\mathrm{FF}}(\theta)$, $\kappa$ and $n$ have been included in the effective coupling parameter $\gamma_{\mathrm{FF}}(\theta) = \beta_{\mathrm{FF}}(\theta) n^2 /  \kappa$.

\begin{figure}[t!]
 \centering
 \includegraphics[width=0.5\textwidth]{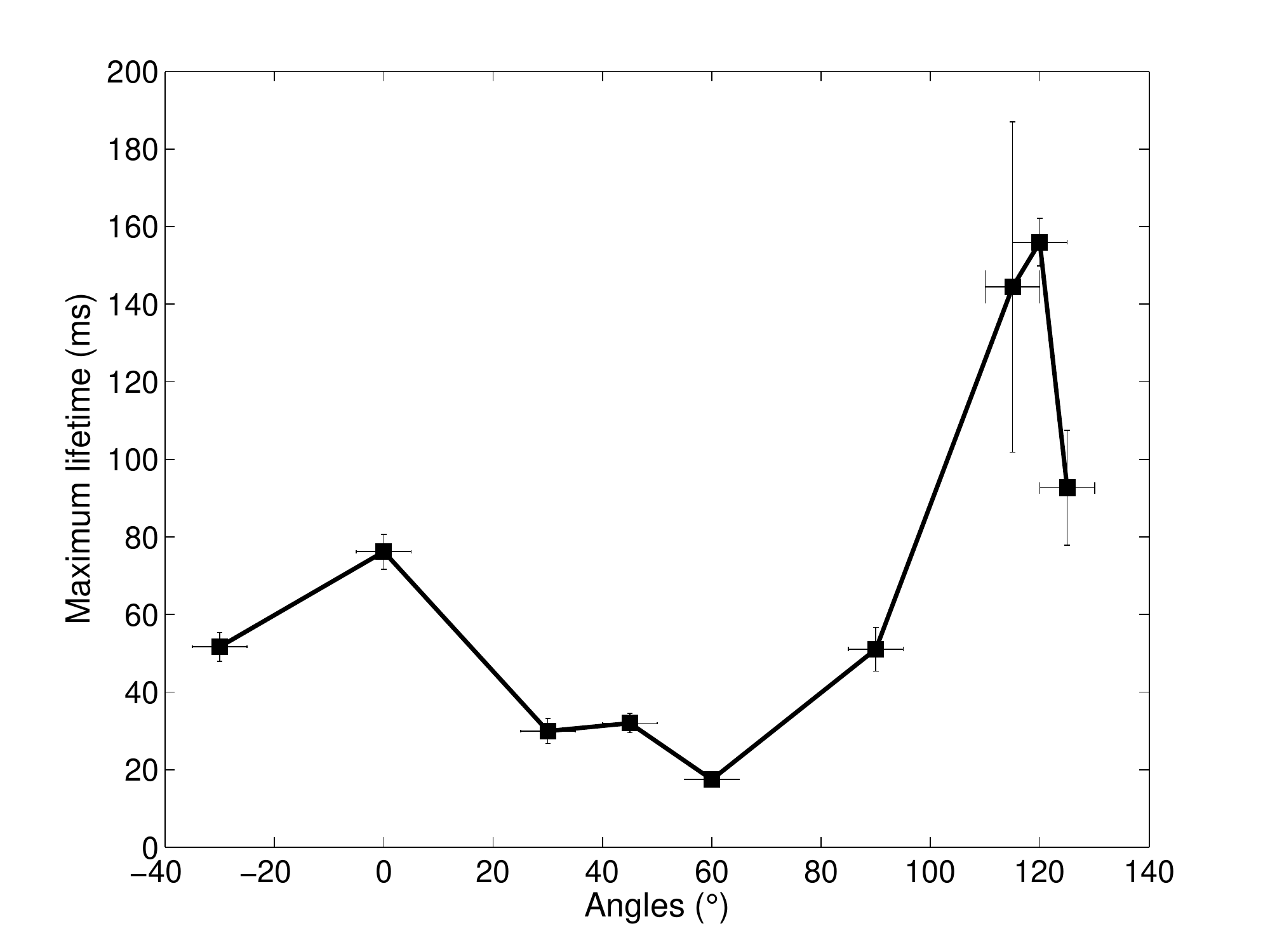}
 \caption{Maximum spectral hole lifetime for each of the measured angles in the crystal with 30 ppm Nd$^{3+}$ concentration at a temperature of 3 K. We can distinguish a global maximum at $\theta = 120\degree$ and a local maximum at $\theta=0\degree$. As discussed in Sec.~\ref{subsec:field}, the global maximum is at $\theta=120\degree$ because the FF rate is minimal at this angle.}
 \label{figure_results3}
\end{figure}

The spectral hole lifetimes measured as a function of magnetic field strength were fitted to the theoretical model using Eqs.~(\ref{eq1}), (\ref{eq3}) and $T_1^{\mathrm{SHB}}=1/(R_{\mathrm{FF}}+R_{\mathrm{SLR}})$. The model yielded a good fit for all eight angles, as shown for three of the angles in Fig.~\ref{figure_results1}. This supports our assumption that there is a linear increase of the spin linewidth as a function of field strength, which causes a reduction of the cross relaxation in the low-field limit. To further strengthen this conclusion it would be highly interesting to measure the spin linewidth as function of strength and angle of the magnetic field, for instance using the methods applied in Ref. \cite{Laplane2016}. Recently, Welinski \textit{et al.} took steps in this direction by measuring the angular dependence of the spin linewidth in erbium-doped Y$_2$SiO$_5$. We also note that different Kramers ions could display different field dependence of the spectral hole lifetime, in the low-field limit, if the field-independent part of the spin linewidth $\Gamma_0$ is larger than $\kappa B$. 

As a result of the competition between the cross relaxation and the SLR direct process there is a maximum spectral hole lifetime for each magnetic field angle (Fig.~\ref{figure_results1}). This maximum has a strong angular dependence and the field strength at which it is reached also depends on the angle. As shown in Fig.~\ref{figure_results3} the maximum lifetime goes from $17.5\pm 1.1~\mathrm{ms}$ at $B=0.55~\mathrm{T}$ for $\theta=60\degree$, to $156\pm 6~\mathrm{ms}$ at $B=0.4~\mathrm{T}$ for $\theta=120\degree$, as shown in Fig.~\ref{figure_results3}. There is also a local maximum of $76\pm 4~\mathrm{ms}$ at $B=0.65~\mathrm{T}$ for an angle of $\theta=0\degree$. In the following we will discuss how the maximum lifetime, and the field at which it is reached, is a complex interplay between the angular variations in the $g$ factor and in the coupling parameters for SLR and FF.

\begin{figure}[t!]
 \centering
 \includegraphics[width=0.5\textwidth]{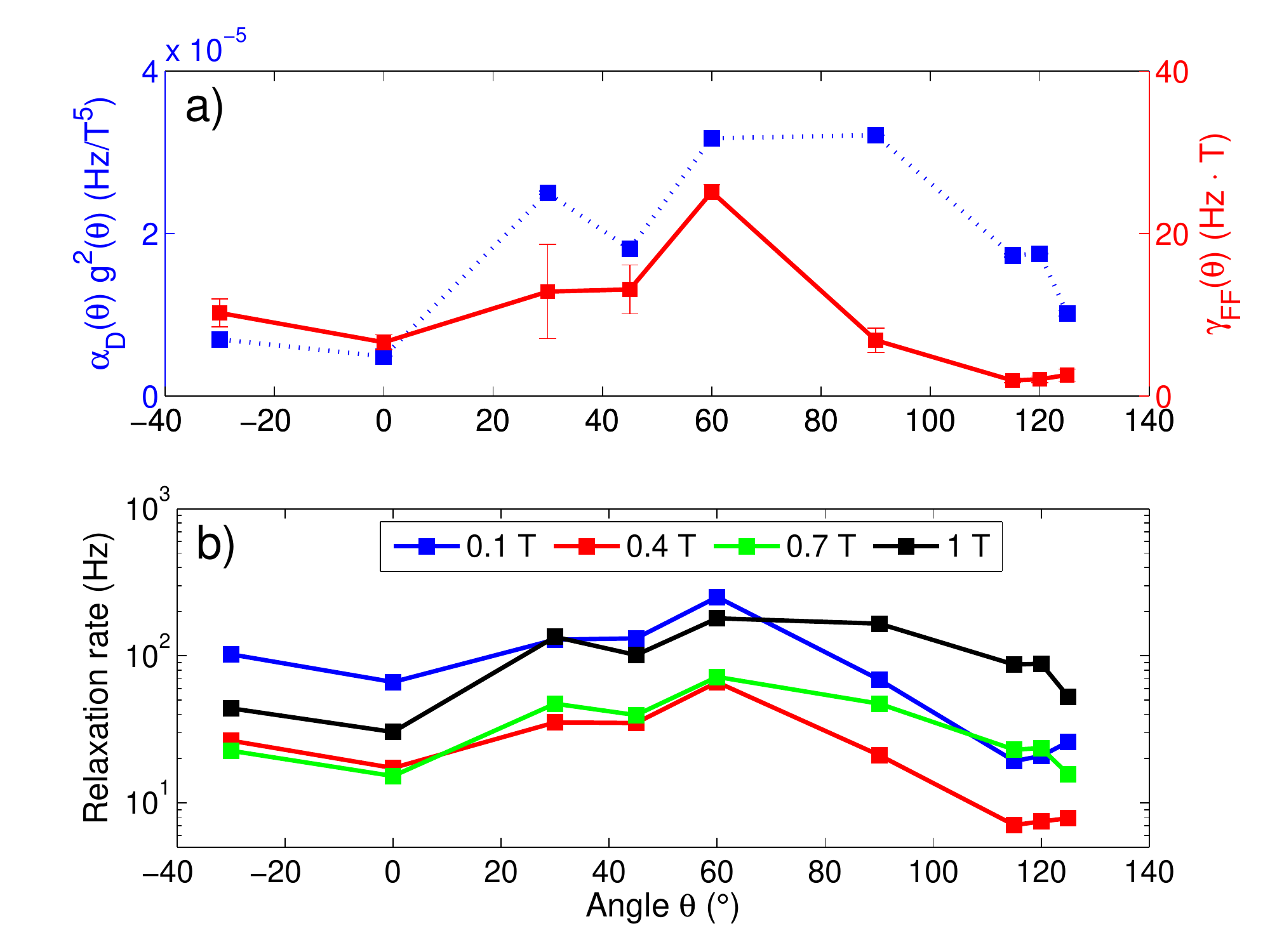}
 \caption{Angular dependence of the coupling parameters presented in the main text and the relaxation rate. In (a) we show the angular variation of $\alpha_{\mathrm{D}}(\theta) g(\theta)^2$ (in units of Hz/T$^5$) and $\gamma_{\mathrm{FF}}(\theta)$ (in units of  $\mathrm{Hz} \cdot \mathrm{T}$). The direct SLR process varies somewhat like a sine function, with a minimum around $\textbf{D}_1$, while the FF appears to have two minima, at $\textbf{D}_1$ and $\theta=120\degree$, respectively. In (b) we show the angular dependence of the total relaxation rate $R_{\mathrm{SLR}} + R_{\mathrm{FF}} $ for $B$ = 0.1, 0.4, 0.7 and 1 T. The minimum rate at about $\theta=120\degree$, for a field of 0.4 T, explains the maximum lifetime of 156 ms observed for this field strength (cf Fig.~\ref{figure_results3}).}
 \label{figure_results2}
\end{figure}

For the cross relaxation process the angular dependence is given by the FF coupling parameter $\gamma_{\mathrm{FF}}(\theta)$. For the SLR direct process we need to consider the product of the coupling parameter and the $g(\theta)^2$ factor, as discussed in Sec.~\ref{subsec:theory_SLR}. For both processes we neglect, for simplicity, the dependence on the ratio $\Delta E (\theta) / (2 k_\mathrm{B} T)$. This is a good approximation for fields of 1 T or less. In Fig.~\ref{figure_results2}(a) we thus show $\alpha_{\mathrm{D}}(\theta) g(\theta)^2\mu_B^5$ and $\gamma_{\mathrm{FF}}(\theta)$. The SLR direct process alone has a minimum rate at $\theta=0\degree$ and maximum rate at $\theta=90\degree$, suggesting that in absence of cross relaxation the optimal field orientation would be along the $\textbf{D}_1$ rather than at $\theta=120\degree$ (see Fig.~\ref{figure_results3}). The cross relaxation has a very different behaviour, with two local minima at $\theta=0\degree$ and $\theta=120\degree$, and a maximum rate at $\theta=60\degree$.

The angular dependencies of the two processes show that there should be two local minima of the total rate, at $\theta=0\degree$ and $\theta=120\degree$, which corresponds well to the observed maxima of the spectral hole lifetime at those angles (see Fig.~\ref{figure_results3}). We also note that the particularly short lifetimes around $\theta = 60\degree$ are due to the fact that both coupling parameters are large in this region. Looking at Fig.~\ref{figure_results2}(a) it is not directly evident, however, why the maximum at $\theta=120\degree$ gives a particularly long lifetime. It is not surprising that it cannot be deduced directly from Fig.~\ref{figure_results2}(a), as both processes have a different dependence on the field strength $B$. To better understand the maximum at $\theta=120\degree$ we plot the total relaxation rate $R_{\mathrm{SLR}} + R_{\mathrm{FF}} $ as a function of angle for several magnetic field strengths, see Fig.~\ref{figure_results2}(b). At the lowest field (0.1 T) the rate is entirely dominated by the FF process and the lowest rate is reached around $\theta=120\degree$. Increasing the field to 0.4 T results in a similar angle dependence, hence the rate is dominated by the FF process, but with significantly lower rate. Since 0.4 T is the field strength at which the longest lifetime of 156 ms is reached, we can conclude that it is given by the low FF coupling parameter at $\theta=120\degree$. Further increasing the field increases the SLR rate and progressively shifts the minimum rate towards $\theta=0\degree$. At the highest field of 1~T it is entirely dominated by the SLR process.

The angular variation of the direct SLR relaxation is defined by the wavefunctions of the Kramers doublets within the $^4\mathrm{I}_{9/2}$ ground state \cite{Larson1966}. Therefore one requires knowledge of the crystal-field Hamiltonian of Nd$^{3+}$:Y$_2$SiO$_5$ in order to make a comparison with our experimental results. To our knowledge, however, the crystal-field Hamiltonian has not been determined. Concerning the angular dependence of the flip-flop process, we have made comparisons with the simple theoretical model discussed in Sec.~\ref{subsec:theory_FF} and further developed in the Supplemental Material. The model predicts a minimum rate in the region around $\theta=110\degree$, in rather good agreement with the data in Fig.~\ref{figure_results2}(a). It completely fails to describe, however, the second minimum at $\theta=0\degree$. A possible explanation could be that the spin linewidth $\Gamma$ has an angular dependence, as recently observed in erbium-doped Y$_2$SiO$_5$ \cite{Welinski2016b}, which is not included in our simple model. To make further progress one would need to experimentally measure the spin linewidth as a function of magnetic field and its orientation in neodymium-doped Y$_2$SiO$_5$.

We conclude this section by emphasizing that optimization of both the magnetic field magnitude and its direction is important when both cross relaxation and SLR play a role in the spectral hole lifetime. The importance of the spin linewidth also suggests that the flip-flop rate could be reduced by co-doping the sample with another rare-earth ion. An increase in the spin linewidth due to co-doping was observed in Er$^{3+}$:Y$_2$SiO$_5$ \cite{Welinski2016b}, by co-doping with Sc$^{3+}$. However, co-doping also results in an increase in the optical inhomogeneous broadening \cite{Boettger2008,Welinski2016b}. It is an open question, then, if it is possible to reduce the spin flip-flop rate significantly by co-doping, without causing a large decrease in the optical depth due to a simultaneous increase in the optical linewidth. 

\subsection{Temperature dependence}
\label{subsec:temp}

\begin{figure}[b]
 \centering
 \includegraphics[scale=0.47]{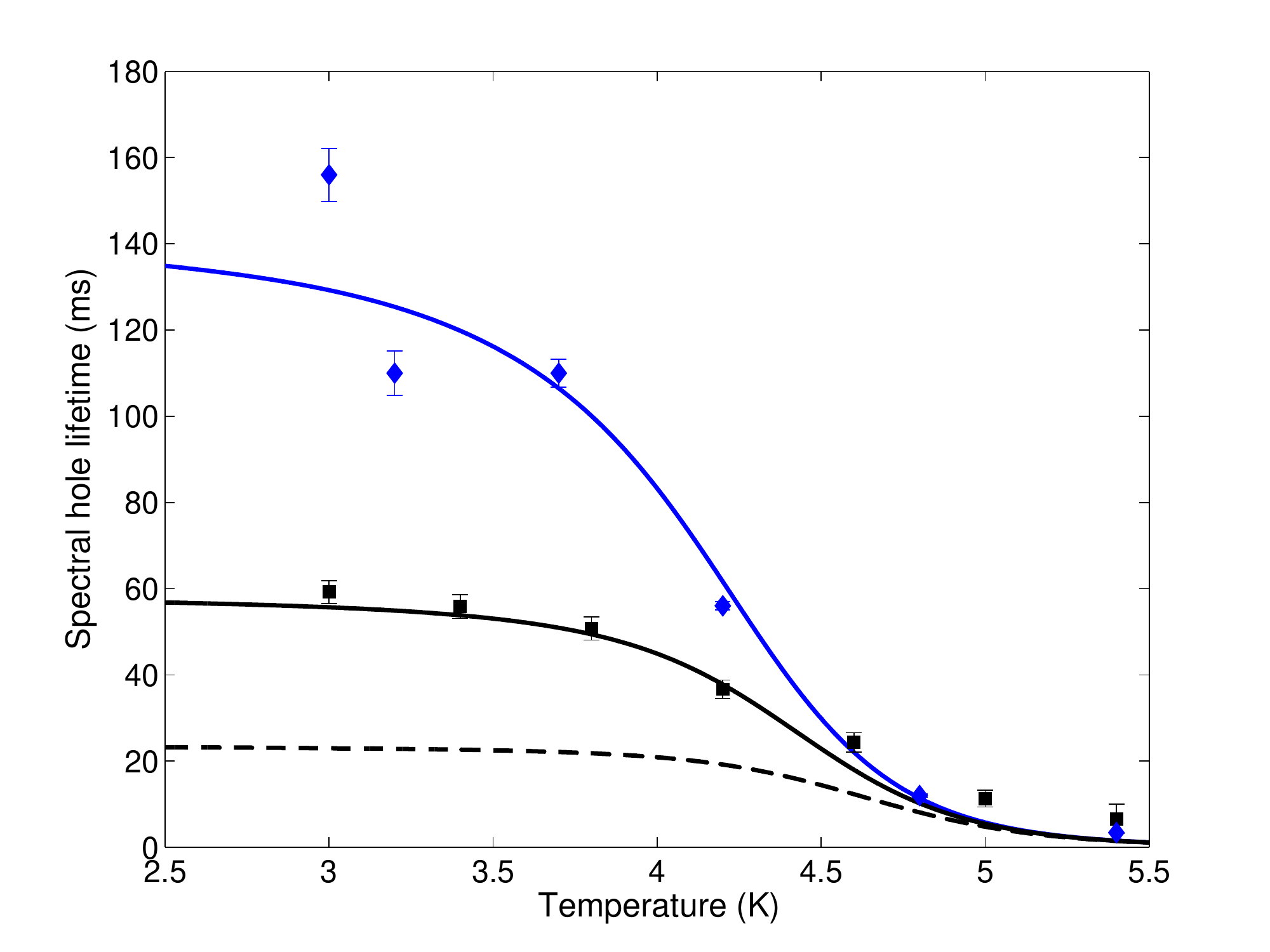}
 \caption{Spectral hole lifetime as a function of temperature for the Y$_2$SiO$_5$ crystals doped with $30\ \mathrm{ppm}$ (blue diamonds) and $75\ \mathrm{ppm}$ (black squares) of Nd$^{3+}$ ions. The magnetic field strength was $B=0.3\ \mathrm{T}$, oriented with an angle of $\theta=120\degree$ with respect to the $\textbf{D}_1$ axis. Both full and dashed lines represent different models used to interpret the data, and are discussed in detail in Sec.~\ref{subsec:temp}.}
  \label{figure_results4}
\end{figure}

We now turn to the temperature dependence of the spectral hole lifetime. We measured the lifetime at different temperatures ranging from 3 to 5.5~K, for the crystals with 30~ppm and 75~ppm  Nd$^{3+}$ concentrations, see Fig.~\ref{figure_results4}. The measurements were made for the optimal angle of $\theta = 120\degree$ (cf. Fig.~\ref{figure_results3}). The static magnetic field in these measurements was given by a permanent magnet installed inside the cryostat, as opposed to the superconducting magnet used for all other measurements. The field was estimated to be 300 mT, but the exact value could be a few tens of mT higher or lower.

At the lowest temperatures the lifetimes in the 30~ppm doped crystal are about twice as long as compared to the 75 ppm doped crystal, which shows that also at the optimal angle the spin FF process is important for these low concentrations. This concentration dependence is a further strong indication that the lifetimes depend on the FF process at this field strength. For both concentrations the lifetime decreases rather slowly at the lowest temperatures, which is due to the linear temperature dependence of the direct SLR process. Above 4 K the lifetimes decrease more rapidly and both samples reach similar lifetimes of less than 20 ms at around 5~K. The rapid decrease is due to the Raman and Orbach processes, which have strong temperature dependence as discussed in Sec.~\ref{subsec:theory_SLR}. It is also expected that the lifetimes at temperatures higher than 5~K do not depend on the Nd$^{3+}$ concentration, since neither the Raman process nor the Orbach process depend on the concentration. 

For the 30 ppm doped crystal we compare the data with the model as fitted to the field-dependent data in Sec.~\ref{subsec:field}, including the Raman and Orbach contributions, with no further tuning of the parameters. As discussed in Sec.~\ref{subsec:theory_SLR} we use the Raman and Orbach parameters measured independently by Kurkin and Chernov \cite{Kurkin1980} using EPR. The agreement with our spectral hole lifetime measurements is rather good (Fig.~\ref{figure_results4}). 

One can now use the model developed for the 30 ppm doped crystal in order to predict the spectral hole lifetimes for the 75 ppm doped crystal. The SLR processes do not have a concentration dependence, while the FF process is expected to have a quadratic dependence $n^2$, see Eq.~(\ref{eq2b}) in Sec.~\ref{subsec:theory_FF}. In Fig.~\ref{figure_results4} we compare the predicted lifetimes with the experimental data by only scaling the flip-flop parameter $\gamma_{\mathrm{FF}}(\theta)$ for the 30~ppm crystal by $(75/30)^2$. This model predicts too short lifetimes at low temperatures, as shown by the dashed line in the graph, which suggests a different concentration dependence. In Fig.~\ref{figure_results4} we also show the prediction based on a linear scaling $(75/30)$ of the $\gamma_{\mathrm{FF}}(\theta)$ parameter, which perfectly reproduces the experimental data. 

The discrepancy with Eq.~(\ref{eq2b}) is possibly due to a concentration dependence of the spin linewidth $\Gamma$. Kittel and Abrahams have shown, for instance, that the dipolar broadening of a spin resonance line depends linearly on the concentration (at low concentrations) \cite{Kittel1953}. It is also possible that the linear coefficient $\kappa$ is concentration dependent. We again emphasize that measurements of the spin linewidth as a function of field strength, field angle and concentration would be highly valuable for understanding the details of the cross relaxation process. To this end, one could do OD-EPR measurements using the method developed in \cite{Laplane2016}.

\subsection{Concentration dependence}
\label{subsec:concentration}

\begin{figure}[h]
 \centering
 \includegraphics[scale=0.47]{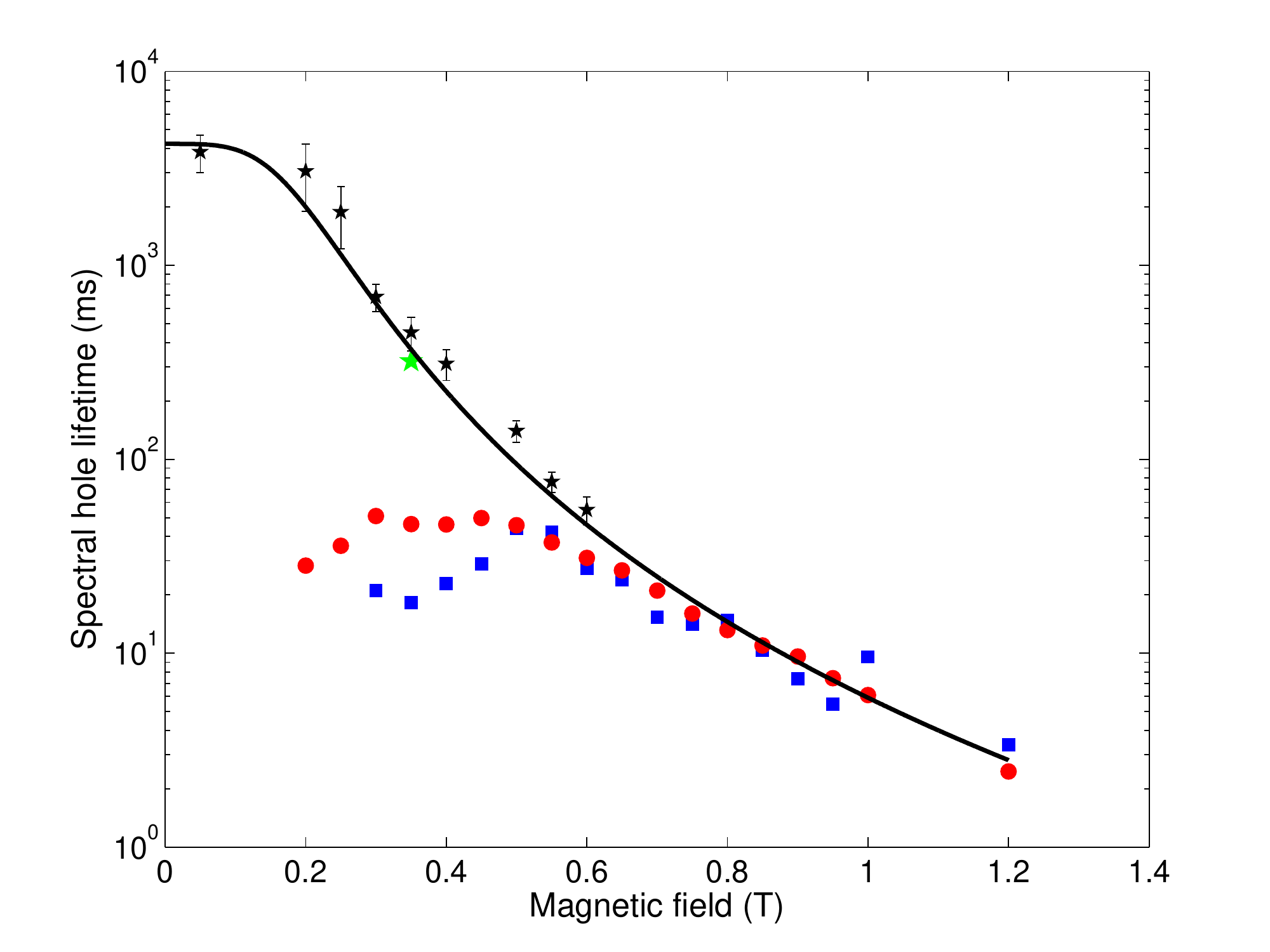}
 \caption{\label{figure7} Spectral hole lifetimes for the crystals with $30\ \mathrm{ppm}$ (red circles), $75\ \mathrm{ppm}$ (blue squares) and $\leq 1$ ppm (black stars) Nd$^{3+}$ doping concentrations. The temperature was 3 K and the field was oriented along the $\textbf{D}_2$ axis ($\theta=90\degree$). We also provide an extrapolation of the model for the spectral hole lifetime (solid line) for a crystal where there is no flip-flop mechanism (see Sec.~\ref{subsec:concentration} for details). The green star shows the lifetime achieved by burning a very large spectral hole in the $30\ \mathrm{ppm}$ crystal, as discussed in Sec.~\ref{sec:PIT}.}
\end{figure}

To further investigate the concentration dependence of the spectral hole lifetimes we compare the lifetimes for three different Nd$^{3+}$ concentrations. In addition to the crystals doped with 30 and 75 ppm of Nd$^{3+}$ ions, we also include a comparison with the Eu$^{3+}$:Y$_2$SiO$_5$ crystal containing a small Nd$^{3+}$ impurity concentration of less than 1 ppm (see Sec.~\ref{subsec:crystal}).

In Fig.~\ref{figure7} we compare the spectral hole lifetimes as a function of magnetic field strength for all three crystals. The field was oriented along the $\textbf{D}_2$ axis ($\theta=90\degree$) and the temperature was 3 K. We emphasize that the measurement data for the 30 and 75 ppm of Nd$^{3+}$ crystals are identical to those shown in Fig.~\ref{figure_results1}. 

The spectral hole lifetimes measured in the crystal doped with $\leq 1$ ppm of Nd$^{3+}$ ions shows a radically different behaviour with respect to the crystals with higher doping. The lifetime increases monotonically as the field strength is reduced and reaches a plateau at low fields. The maximum lifetime is 3.8$\pm$0.8 s, which is a 75-fold increase with respect to the maximum lifetime of 51$\pm$6 ms obtained in the 30 ppm  Nd$^{3+}$ doped crystal, for this orientation of the field (cf. Fig.~\ref{figure_results3}). Clearly the cross-relaxation process at low magnetic fields poses a serious limitation on the maximum achievable lifetime, even at doping concentrations as low as 30 ppm. Although the extremely low-doped crystal results in long-lived spectral holes, the associated low optical depth prevents it from being used directly in quantum memory applications, as it would lead to very low storage efficiencies. A potential solution is to use cavity-enhanced quantum memory schemes \cite{Moiseev2010a,Afzelius2010a}, or a slightly more doped sample, or even a combination of both these approaches.

In Fig.~\ref{figure7} we also show an extrapolation of the model fitted to the field-dependent data obtained for the crystal doped with 30 ppm of Nd$^{3+}$ ions. Specifically we use the fitted $\alpha_{\mathrm{D}}(\theta)$ parameter at $\theta=90\degree$ to calculate the direct process, to which we add the Raman and Orbach rate contributions. The flip-flop contribution is not included, such that the spectral hole lifetime is calculated using only $T_1^{\mathrm{SHB}}=1/R_{\mathrm{SLR}}$ and Eq.~(\ref{eq1}). The agreement with the lifetimes measured in the extremely low-doped sample is excellent, for all measured field strengths. Note that the signal-to-noise ratio was too low to measure the lifetime beyond 0.6 T, due to the low optical depth. 

We conclude that the flip-flop process has been completely eliminated at a concentration of $\leq 1$ ppm of Nd$^{3+}$ ions. We estimate that the already observed linear dependence of the lifetime on concentration can in itself explain this effect (see Sec.~\ref{subsec:temp}). In addition one could expect that the presence  of Eu$^{3+}$ ions further reduces the flip-flop rate by increasing the Nd$^{3+}$ spin linewidth.

\subsection{Lifetime of a spectrally large hole}
\label{sec:PIT}

In the previous sections we have seen strong evidence for a concentration dependence of the spectral hole lifetimes due to cross relaxation. As explained in Sec.~\ref{subsec:theory_FF} this decay process is possible because only a small spectral region is spin polarized during the pumping process (ensemble A), while the majority of spins (typically 99\% or more) are not affected by the pumping process (ensemble B). As a consequence one would expect that the spectral hole lifetime would change if a very large spectral hole was burnt into the optical linewidth, cf. Fig.~\ref{figure2}, which would largely reduce the number of spins in ensemble B and increase the spins in ensemble A. The result should be a strongly reduced spin flip-flop probability and long spectral hole lifetime.

We investigated this possibility in the crystal doped with 30 ppm of Nd$^{3+}$ ions. The magnetic field strength was 350 mT, oriented along $\textbf{D}_2$ ($\theta=90\degree$), and the temperature was 3 K. It is important to note that a minimum magnetic field strength is required to be able to split the ground state doublet more than the optical inhomogeneous broadening, as illustrated in Fig.~\ref{figure2}. This is a condition in order to be able to optically spin polarize a large part of the inhomogeneous spectrum. This prevents the technique to be used at very low fields, where the flip-flop process dominates, unless the inhomogeneous broadening is particularly weak.

\begin{figure}
 \centering
 \includegraphics[width=0.5\textwidth]{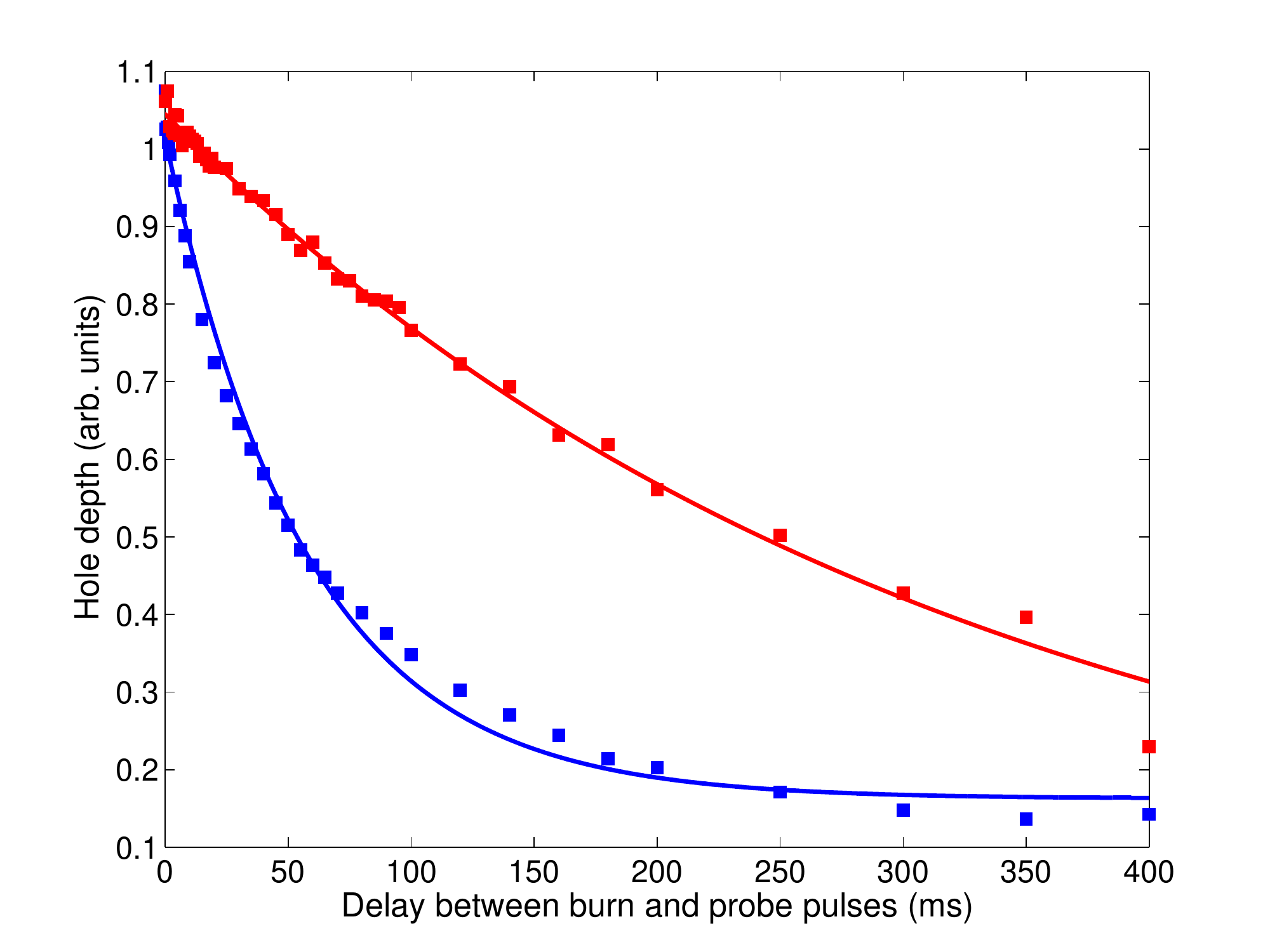}
 \caption{\label{figure8} Spectral hole decay measurements for a narrow spectral hole (blue squares) and a large spectral hole (red diamonds). The narrow hole has a linewidth of a few MHZ, while the large hole represents optical pumping of about 50\% of the ions within the optical inhomogeneous linewidth of about 6 GHz. The solid lines show the fitted single-exponential curves, which give lifetimes of 50 ms (blue line) and 320 ms (red line), respectively.}
\end{figure}

In Fig.~\ref{figure8} we show the decay curve of both a narrow hole and a large and deep spectral hole where about half of the inhomogeneous broadening has been optically pumped. The spectral hole lifetime increases from 50 ms to 320 ms by burning a large hole, a rather spectacular increase. If we compare this lifetime with the one obtained for the extremely low-doped crystal it actually gives a similar lifetime, as shown by the green star in Fig.~\ref{figure7}. This is rather surprising since we estimate that only about 50\% of the spins where polarized by burning a large hole.

While this experiment shows the potential of modifying the cross relaxation rate by optical pumping, the detailed underlying mechanism is not well understood and further experiments will be necessary to be able to apply this method to various quantum memory schemes. Another interesting perspective is to decrease spectral diffusion using optical pumping of large spectral regions. Indeed, in many cases the optical coherence times of Kramers ions are limited by spectral diffusion driven by spin flip-flops at low fields \cite{Bottger2006a}.

A related and interesting question is if the cross-relaxation rate can be suppressed by decreasing the temperature such that all spins point in the same direction. This limit is reached when $\Delta E (\theta) \ll (2 k_\mathrm{B} T)$ where all the spins would polarize into the lower Zeeman state (cf.~Fig.~\ref{figure2}) by the low temperature. Looking at Eq.~(\ref{eq2b}) one would naively conclude that the cross relaxation rate would be highly suppressed. However, if a small fraction of the spins are excited to the higher Zeeman state, then rapid spin flip-flops will occur with the spins in the lower Zeeman state. As the SLR rate should be very weak at such temperatures, we expect the flip-flop rate to be the limiting process for storage of quantum information. Similarly any narrow spectral features created through SHB using these Zeeman states would also decay due to flip-flop rates. But as our results show one could go to much lower concentration to mitigate this problem, or possibly co-dope the material in order to increase the spin linewidth and reduce the cross relaxation rate. We also emphasize that the overall population difference between the two states will of course not be affected by cross relaxation, which is the typical measure of spin population lifetimes at extremely low temperatures (tens of mK) \cite{Probst2013}. Hence, some caution has to be exercised when using such measurements to predict useful spin coherence times, which are likely to be affected by spin flip-flops, rather than SLR processes, at low temperatures.

\section{CONCLUSIONS AND OUTLOOK}
\label{sec:Conclusions_Outlook}

We have studied the relaxation mechanisms of spectral holes in neodymium-doped orthosilicate under different magnetic fields, temperatures and dopant concentrations. Our main finding is that the limiting factor in achieving long-lived spectral holes is the spin cross relaxation, or flip-flop, process. We have also shown that both the strength and angle of the magnetic field must be carefully optimized to maximize the hole lifetime. By decreasing the concentration to as low as $\leq 1$ ppm of Nd$^{3+}$ ions, we could eliminate the flip-flop process and reach a hole lifetime of 3.8 s at 3 K. Our results show that optimization of the dopant concentration is crucial in order to find a compromise between the spectral hole lifetime and optical absorption coefficient. We have also shown that the cross relaxation rate can be drastically reduced by creating very large spectral holes, which could open up new ways of engineering spectral hole lifetimes in a given crystal and to improve optical pumping in crystals doped with rare-earth Kramers ions. We also argue that lowering the temperature to the mK regime would not suppress the effect of cross relaxation on quantum coherence measurements. We believe that these results should allow better optimization of these crystals for applications in quantum memories \cite{Bussieres2013,RiedmattenAfzeliusChapter2015} and narrow-band spectral filtering for biological tissue imaging \cite{Li2008,McAuslan2012a,Zhang2012}.

\section{ACKNOWLEDGEMENTS}

We thank Florian Fr\"{o}wis for useful discussions and Pierre Jobez for experimental support. This work was financially supported by the European Research Council (ERC-AG MEC) and the National Swiss Science Foundation (SNSF) (including Grant No. PP00P2-150579). J. Lavoie was partially supported by the Natural Sciences and Engineering Research Council of Canada (NSERC).

\bibliography{Main.bbl}

\end{document}